# Pushing Software-Defined Blockchain Components onto Edge Hosts

Mayra Samaniego
University of Saskatchewan
mayra.samaniego@usask.ca

Ralph Deters
University of Saskatchewan
deters@cs.usask.ca

## Abstract

*With the advent of blockchain technology, some management tasks of IoT networks can be moved from central systems to distributed validation authorities. Cloud-centric blockchain implementations for IoT have shown satisfactory performance. However, some features of blockchain are not necessary for IoT. For instance, a competitive consensus. This research presents the idea of customizing and encapsulating the features of blockchain into software-defined components to host them on edge devices. Thus, blockchain resources can be provisioned by edge devices (e-miners) working together closer to the things layer in a cooperative manner. This research uses Edison SoC as e-miners to test the software-defined blockchain components.*

## 1. Introduction

Internet of Things (IoT) networks enable connectivity with the real world anytime and anywhere [1]. The pervasiveness characteristic of IoT networks would make IoT devices (aka things [2]) the primary enablers of data [3].

IoT networks have experimented exponential scalability [3]. This scalability introduces management challenges at the constrained network. For instance, verification of identities and correct configurations of the resources of the things network, provenance data to analyze the behaviour of the things network, and validation of transactions from the things and over them.

Some studies have designed blockchain-based implementations to handle some of these challenges. For instance, Kaku et al. [4] present a blockchain-based system to handle provenance of responses in IoT. This kind of implementations is considered a cloud-centric one [5], in which there is a service layer that connects the things network and the services hosted in the cloud.

Cloud-centric blockchain implementations would not involve any challenge because cloud computing provides access to virtualized and scalable services over the Internet [6]. Besides, even though the cloud represents a robust and reliable architecture for IoT analytics, its consolidated power might not fit the dynamic characteristics of IoT networks [7]. These systems introduce significant latency, network traffic and bandwidth consumption [8].

Fog computing extends the cloud features toward the edge of IoT networks to deal with specific characteristics of some networking scenarios such as a broad set of heterogeneous nodes, geographical location, and real-time communication [6] [9]. Cisco explains that fog nodes can directly access physical IoT devices, consequently reducing latency and bandwidth consumption [8]. According to Bonomi et al. [10], IoT analytic tasks can be moved to a fog network as well. In our most recent work, we implemented a blockchain-based middleware called Amatista [11]. Amatista implements zero-trust hierarchical management in a blockchain-based fashion.

This research proposes going to a deeper level, the edge level [12], and separate the features of blockchain to encapsulate them into software-defined components [13][14]. Thus, the edge network can host these software-defined blockchain components. This approach eliminates the dependency on either the cloud or fog network. This approach breaks the barrier of the constrained computing capabilities because each component can be hosted on a different physical device but working collaboratively to build a blockchain system.

The rest of the paper is organized as follows. Section 2 studies blockchain in IoT. Section 3 introduces software-defined components in IoT, and the architecture proposed by this research. Section 4 presents the experiments and evaluations. Finally, section 5 presents the conclusions of this research.

## 2. Blockchain & IoT

### 2.1. Blockchain

Blockchain protocols started attracting the attention of researchers in 2009 when Satoshi Nakamoto (an online pseudonym) introduced the Bitcoin cryptocurrency system [15]. The blockchain that supports Bitcoin cryptocurrency is a public peer-to-peer distributed ledger that records all transactions within the Bitcoin network. This is the characteristic of distributed in-chain database and synchronization across the network.

Bitcoin implements a public blockchain network, which is open to any participant on the Internet [15]. Participants that are part of the Bitcoin network are called miners because their task is mining blocks of transactions to be written in the chain.

Mining is the name that the process of verifying blocks of transactions in Bitcoin receives [15]. In Bitcoin, the consensus

H I C S S

mechanism that the network uses for mining is called proof of work. The proof-of-work mechanism states that a miner must solve a cryptographic problem to gain the right of writing the block to the chain. Miners must proof trustworthiness solving the cryptographic problem, which requires high computing power.

Other consensus algorithms do not require that participants spend high computing power. For instance, Practical Byzantine Fault Tolerance (PBFT) algorithm [16] or Round Robin (RR) scheduling [17].

The consensus is the feature of blockchain that prevents the execution of incorrect or unreliable transactions. This feature encourages reliability between nodes that do not know each other [18].

Unlike Bitcoin, there are also private blockchain networks. Private blockchain networks limit the access of participants and the execution of transactions [19]. The way a private blockchain network validates and accepts participants and transactions may vary, depending on internal rules or technology [20]. Ripple [21] and Hyperledger Fabric [22] are examples of private blockchain networks.

Blockchain protocols do not allow changes in the transactions that a mined block contains [23]. This is the characteristic of immutability. This approach encourages data sharing between parties that do not trust each other.

Some blockchain technologies allow storing business rules (programming functions) in-chain. Thus, we can define more complex interactions between parties. This feature is generally known as smart contracts [24]. Smart contracts is the name that Ethereum uses to call this in-chain programmed functions [25]. In this research, we have adopted the name of smart contracts as well. Depending on the blockchain technology, smart contracts might receive other names. For instance, Hyperledger Fabric calls it in-chain code [22]. Additionally, each blockchain technology might have their protocol and different programming languages to build and deploy smart contracts in the network, in case they support them.

In blockchain-based implementations, transactions between parties are stored and validated in a distributed manner without depending on a central validation authority [15].

## 2.2. Blockchain for IoT

In blockchain systems with a cryptocurrency, like Bitcoin [15], miners compete between each other to gain the cryptocurrency reward after mining a block. This approach encourages miners to invest resources in the blockchain network, but at the same time, it makes the miners behave individually.

IoT networks do not have the computing power to make miners compete for [26][27]. This is the main reason why blockchain implementations for IoT are mostly hosted in the cloud. For instance, Sharma et al. [28] present a cloud-centric blockchain architecture to address data-related issues in IoT, such as availability and delivery.

Additionally, Dorri et al. [29] present a blockchain-based framework to handle security and privacy in IoT. However, in this solution, miners work as central validation authorities. There is no consensus achieved between miners.

Implementing blockchain in the cloud means that all the blockchain features are executed together by powerful nodes [30]. The cloud introduces the benefits of efficient use and orchestration of resources, on-demand self-service and rapid deployment, and elasticity [30]. However, these benefits introduce high costs. It would be expensive to store every single IoT transaction in a cloud blockchain.

Some works have implemented blockchain towards fog networks to deal with the lack of engagement over the things that cloud blockchain implementations present. For instance, in our previous works, we designed fog solutions that store virtual resource configuration [31], enhance the fog network with some artificial intelligence features [32] and provide multi-tenancy [33].

Even though, blockchain protocols have contributed to IoT management, until now we have depended on either the cloud or the fog to implement them. Again, this is mainly because traditional blockchain implementations make miners compete between each other, which requires high computing power. High computing power is not part of the IoT paradigm [1].

Nowadays, edge computing allows applications to execute some processing tasks closer to the things network [12]. This approach would contribute to fulfilling the things-oriented vision of IoT [34]. The things-oriented vision of IoT states that the enhancement of the things is the priority. However, the limited computing power of edge devices makes it impossible to implement an entire blockchain node in one single physical edge device. That is why it becomes necessary to separate the features of blockchain and encapsulate them into components that can be hosted by edge nodes physically separated but cooperatively committed.

This research seeks to develop a private blockchain implementation for IoT that does not require a cryptocurrency mechanism reward and distributes blockchain features towards the edge network to provide similar functionalities to the ones that cloud/fog-centric blockchain provides but at the edge level. This research visualizes blockchain as a virtual system that can be encapsulated into software-defined components. Thus, there is no limitation of physical computing capabilities.

## 3.   Software-defined blockchain architecture

Software-defined concepts were initially developed to customize virtual Internet networks [13][14][35] and manage network functionalities [36][37].

Similarly, we could use this concept for IoT networks. According to Nastic et al. [38], "Software-defined IoT units are used to encapsulate the IoT resources and lower level functionality in the IoT cloud and abstract their provisioning and governance, at runtime."

This research builds software-defined components of specific features of blockchain to deploy them at the edge of IoT networks. According to Biswas et al. [39], IoT networks would benefit from software-defined ecosystems or virtual systems.

The primary barrier to implement a blockchain technology at the edge of IoT networks is the lack of computing capabilities [1]. This research aims that it is not necessary to implement an entire blockchain technology at the edge of IoT networks, but only those features of blockchain that are necessary to guarantee a reliable operation of the things.

Additionally, this research aims that it is not necessary to implement all the features of blockchain in one single edge node, but the encapsulation of specific features towards different edge nodes. Thus, we can get a distributed software-defined blockchain system that is not limited by physical devices and is not dependent on either the cloud or the fog network.

This research breaks the limitations of computing capabilities by building software-defined blockchain components that are independent enough to be hosted in separated edge devices and at the same time cooperative enough to build a virtual blockchain system.

### 3.1. Software-defined blockchain components hosted towards edge nodes

Deploying software-defined blockchain components onto edge devices introduces the following benefits.

The edge network gains autonomy. There would not be necessary to have a bridge layer that connects the things network with the cloud, because the edge network would handle some management tasks independently, such as the verification of transactions, provenance, and access control.

Reduced latency. As edge nodes would implement management tasks closer to the things network, bandwidth consumption would decrease, and the performance of the IoT infrastructure would improve. Moreover, the most important is that this improvement would not be because the management tasks are moved to the cloud but distributed across the edge of the IoT network.

Time-efficient management. Different edge nodes would execute operations over different segments of things, which would optimize the management time. For instance, updating the configuration of a group of things and updating the access policies of another group of things would be handled by edge nodes separately. This approach shifts the focus from a static data-centric IoT to a dynamic resource-centric (things-oriented vision of IoT [34]).

Independent authorities. An edge node does not necessarily have to execute all the blockchain components hosted on it over the same group of things all the time. For instance, an edge node can participate in the consensus of a specific group of things, and it can execute smart contracts that affect an entirely different group of things.

Provenance. All the transactions from and over the things layer can be tracked.

No computing capabilities limitation. From the perspective of this work, this is the most important benefit of building software-defined blockchain components. As we encapsulate the blockchain features into separated components, they can be distributed towards different edge nodes. Moreover, any edge node can decide to terminate any of these components at any time.

### 3.2. Challenges when implementing software-defined blockchain components towards edge nodes

Implementing software-defined blockchain components at the edge level introduces some challenges.

First, the size of the transaction must be limited to the maximum size admitted by the communication protocol between the things and the edge network. Second, as a cryptocurrency is not necessary, there could be a lack of incentive for edge nodes to validate transactions accurately. Finally, edge nodes do not have enough computing storage to store a large chain of blocks.

This research implements the following actions to handle the challenges mentioned above respectively. First, sensors build small transactions that can be sent faster across the network. Second, instead of having a crypto-currency reward mechanism, there is a provenance reputation mechanism. Finally, edge nodes only keep a chain of metadata of the last n blocks in memory and store these n block files locally.

### 3.3. Architecture

This work calls the edge nodes that host software-defined blockchain components e-miners.

Figure 1 shows the architecture that this research proposes. The first layer represents the network of sensors and actuators at different scenarios.

The second layer represents the e-miners that host the software-defined blockchain components. We represent e-miners as turtles because even though they have limited computing power, they still can show a good performance executing the software-defined components saved in their shell.

This work builds three software-defined blockchain components, smart contracts, consensus, and in-chain data storage. Each e-miner can implement any of these software-defined blockchain components.

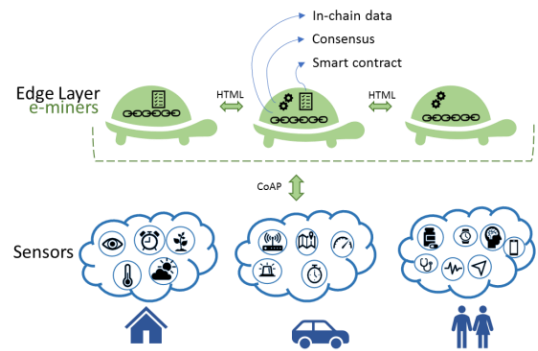

**Figure 1. General architecture of software-defined blockchain components towards edge hosts.**

### 3.3.1. Smart contracts

Traditional smart contracts in the form of Distributed Applications (DApps) in Ethereum [40] distribute a copy of the smart contracts to all participants in the network. Besides, participants can not terminate the smart contract unless there is an ending function that starts working when a condition is met.

The large size of IoT networks makes it necessary to segment the execution of smart contracts. This research introduces a publish/subscribe policy that allows e-miners to select what smart contract to execute. Additionally, because of the energy limitation of e-miners, this research allows them to terminate the execution of a smart contract if that smart contract is being executed by other e-miners, depending on the specific context.

### 3.3.2. Consensus

Traditional consensus mechanisms like the proof of work of Bitcoin [15] or the proof of stake of Ethereum [25] require that all the nodes in the network participate in the consensus process.

In IoT, every node mining every block and verifying every single transaction would cause the network to collapse. We introduce a publish/subscribe policy that allows e-miners to participate in the consensus of blocks that come from specific observed things or other e-miners.

Additionally, in IoT, the primary goal of e-miners would not be to compete to write a block in the chain to get the cryptocurrency reward but ensure that each block of transactions is a valid one before executing and writing it.

According to Tschorsch et al. [41], blockchain can work satisfactorily without having a cryptocurrency. However, the lack of cryptocurrency would also mean the lack of motivation for e-miners.

For this research, we have implemented a practical Byzantine fault tolerance algorithm (PBFT) [16]. Instead of receiving a cryptocurrency reward, e-miners receive a score reward for building their provenance reputation. Also, the e-miner leader that wrote the block must wait a specific time to send another block (round robins algorithm [42]). Thus, the other e-miners can be leaders as well. Each node is responsible for keeping a high reputation to continue being trustable to write blocks in the chain.

### 3.3.3. In-chain data storage

Traditional blockchain data storing requires that all the information of transactions is written in the chain.

In IoT, e-miners would not be able to synchronize the entire chain because of the limited storage capabilities.

Ripple [21] is an example of a blockchain that modifies the in-chain storage. Ripple does not store the provenance of transactions in-chain but only the previous and the new balance with no chargebacks, which makes the network more efficient and scalable.

Following the Ripple's example, this research implements in-chain metadata storage. E-miners store blocks of metadata of an n number of previous blocks mined. Also, e-miners store the files of those n previous blocks mined.

Because of the limited computing capabilities of e-miners, there is the need to have a distributed repository close to them. E-miners discharge the in-chain metadata and local files to a fog repository regularly.

This research designs software-defined blockchain components as independent artifacts. We can customize and update them at any time separately. For instance, we can implement new smart contracts without affecting neither the consensus nor the in-chain data components. In the same manner, we can implement a new consensus mechanism without affecting neither the smart contracts nor the in-chain data storage.

### 3.3.4. Technology and communication

We have used the Constrained Application Protocol (CoAP) [43] to communicate the e-miners with the things network. CoAP was designed to run on devices with limited memory. Additionally, CoAP runs on UDP by default, which saves bandwidth. CoAP follows a REST [33] approach and supports GET, POST, PUT, and DELETE operations. This approach facilitates the direct engagement of things when executing smart contracts over them.

We have used HTTP protocol to handle the communication between e-miners. The size of the block of transactions that e-miners build is too big to use CoAP. However, e-miners implement a CoAP interface to receive the transactions from the things network.

We designed the software-defined blockchain components as full state resources following REST using Go Language [44].

## 4. Evaluations

This section evaluates the performance of e-miners running software-defined blockchain components. Figure 2 shows the architecture for the experiments. The architecture involves two layers. The first layer represents the simulated sensors. This layer is formed by one Intel Edison System on a Chip (SoC) [45] (table 1) plugged on a Spark module [46]. The second layer represents the e-miners. Three Edison SoC's plugged on Arduino boards form this layer. The three e-miners host the in-chain data and consensus software-defined blockchain components. The first e-miner (red turtle) also hosts the smart contract software-defined blockchain component.

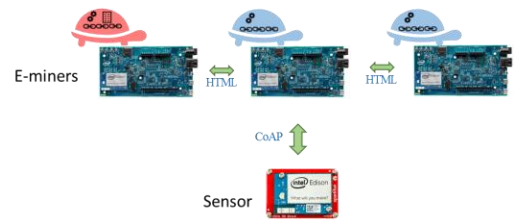

**Figure 2. The architecture for experiments.**

**Table 1. Specifications of Edison SoC [45]**

| Edison System on a Chip | |
|---|---|
| *Operating System* | Linux Yocto |
| *CPU* | 500 MHz dual-core, dual threaded Intel Atom and a 100 MHz 32-bit Intel Quark microcontroller |
| *RAM* | 4GB LPDDR2 SDRAM |

All the software-defined blockchain components have been evaluated under three different intervals between transactions from sensors, 50, 100 and 200 milliseconds. The sensors send one thousand transactions.

## 4.1. Evaluation of smart contract software-defined component

For this experiment, we have designed a smart contract that observes the sensor and analyzes each transaction that it sends. The sensor and the e-miner communicate through the Constrained Application Protocol (CoAP [43]).

Figure 3 shows an example of a transaction that a sensor sends. The transaction has three fields, the hash of the message, the message, and the signature. The hash of the message is generated using SHA-1 [47] algorithm. The sensor encrypts the message and signs the transaction using RSA [48] public-key cryptosystem with a key size of 2048 bits.

The encrypted message contains the raw data in JSON format. The structure of the row data has two fields, the temperature value (v) and the measurement symbol (c, Celsius). The size of the entire transaction is 1099 bytes. The size might vary depending on the value of the temperature field.

```
{
  "Hash":"a7745f817a768e532ba4f98b6827017147a20d11",
  "Msg":"72d91fbbe10faa37e395bcddadif98f769561c7b242a36933316ee7b412512f3cbca1436bad75c32e2e3c895fd1d8bed66872f389b
  067565eb59fde7b46684aed324e0c98f5854c0c5b3c38b79c79725e625451bae73deda99dfcf903445976154c88b83d79fc70cf16f
  54c6567d1647f3e774d69b74da4097bbe59b60bd768b85ef64f35fad595a1d44a12801b0fd9fdb759fbed91206863aa7de6b21c
  eef76a778b591416d0d11d69ce64be6efca936ae9f829718bef4af1b72bd1f1d353215360deeb166e348b7f858cf6e037e26803
  bbf7f9db53af6ecb7f521d92f00609cd08bca81a1c2c4d92b8ca10272fd66 c50f4652af40a89f87612edba7813a4d87",
  "Signature":"5e59797384addf630040b39f3c7bdf99a6fd74ea3f989dc19c15ecf4a08b52da96c37e22726af8549ebd8a6f9a10dc7
  126b55f539b44e46913762396b01366a0a3e4ea9f66edd3 4b853b547d7c9030cc08031bccb6458cc2f7cca54a932e6e34cb33a3c28e106
  3a73aad18aa75f8681c291fb75c76f2c7083bf59493c43c9e2ca3b3519f9b6 9d803305a9a7e74bb5a8dc4be6f72977301356133
  6844406453c0028d3cad3681949e37786ac2c3e0ff24d06a3af56a4cf9a19fe39f949ce29dfc64d5c85ac49c5525e347f7c9182a5
  8f1c1455baad7ad59ec47efde16ab56ef7d62266c3de2ba89cd9dc7b25d683bbc72009ddd422abb0b1c9a4d86aa69c8afad"
}
```

→ {"v":211, "m":"c"}

**Figure 3. Example of a transaction that a sensor sends.**

When the e-miner receives a transaction, it executes the following tasks. First, the e-miner verifies the signature of the transaction. Second, the e-miner decrypts the message, generates the hash value of the decrypted message again, and compares both hashes to verify that the message has not been altered. Finally, the e-miner calls the smart contract that analyzes the raw data. The smart contract triggers an alarm each time the received data meets a pre-defined condition.

Figures 4, 5, and 6 present the results of the evaluation of the smart contract software-defined component under the three intervals of transactions, 50, 100, and 200 milliseconds

respectively. The x-axis represents the number of transactions that the sensor sends, which is one thousand. The y-axis represents the time in milliseconds that the e-miner takes to execute the smart contract component and trigger the event if necessary.

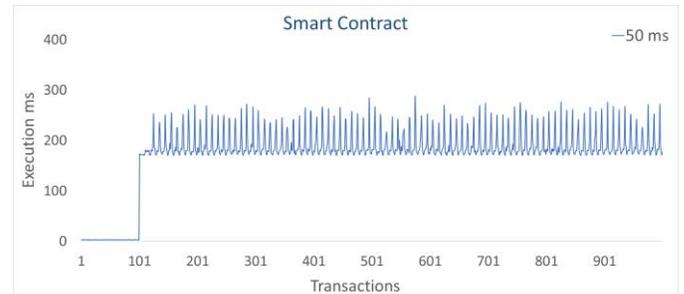

**Figure 4. Evaluation of smart contract. Interval of transactions, 50ms.**

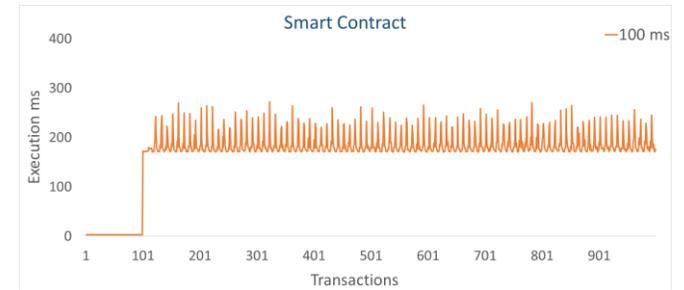

**Figure 5. Evaluation of smart contract. Interval of transactions, 100ms.**

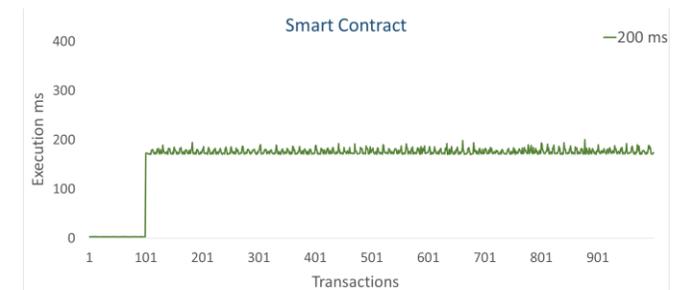

**Figure 6. Evaluation of smart contract. Interval of transactions, 200ms.**

As can be seen in figures 4, 5, and 6, the response patterns for the three intervals of transactions is very similar.

The response times keep consistently low during the first 100 transactions. The e-miner takes 2.8 milliseconds on average to execute the smart contract during the first 100 transactions. However, after the 100[th] transaction, the response times rise from 2.8 milliseconds to 172 milliseconds when the interval between transactions is 50 milliseconds, 168 milliseconds when the interval between transactions is 100 milliseconds, and 156 milliseconds when the interval between transactions is 200 milliseconds.

This behaviour is due to the configuration of the in-chain data component and file storage. For this experiment, we configured the in-chain data component and the local file storage to work after every 100 consensus achievements. After the e-miner receives consensus approval of 100 blocks, the in-chain data component starts building one in-chain block of metadata for each validated block. Additionally, a file for each block is written locally.

If we had configured the two components to start working at a different number of consensus achievements, the increase in the response time would have happened at that specific number. The in-chain component that the e-miner keeps in memory and the files that the e-miner stores locally make the response times increase.

The decrease in the average response times during the first 100 transactions (172, 168, and 156 milliseconds) shows that the bigger the interval between transactions is, the faster response from the e-miner becomes. The delay intervals in which the sensor sends transactions affects the performance of the e-miner executing the smart contract software-defined component.

The graphs also show some peaks in all the intervals of transactions. This is due to the block fabric activation, the erasing process of the in-chain data component, and the erasing of the local file storage. After the e-miner receives 10 transactions, it builds the block that contains those 10 transactions and propagates it for consensus. Additionally, when the in-chain component has 10 blocks, the e-miner erases the in-chain storage and moves the local files to a permanent Fog storage.

These picks increase by 9 points on average to reach response times of up to 286 milliseconds when the interval between transactions is 50 milliseconds, 8 points on average to reach response times of up to 272 milliseconds when the interval between transactions is 100 milliseconds, and 7 points on average to reach response times of up to 200 milliseconds when the interval between transactions is 200 milliseconds. The bigger the interval between transactions is, the lower the peaks reach.

## 4.2. Evaluation of consensus software-defined component

For this experiment, we have implemented the Practical Byzantine Fault tolerance algorithm (PBFT) [16] between the three e-miners. After an e-miner finishes mining 10 transactions, it builds a block and propagates it to the other e-miners to get consensus approval.

Figure 7 shows an example of the block that the e-miner builds. The block has four fields, previous hash, data hash, message, and signature.

The e-miner generates the hashes using SHA-1 [47] algorithm.

The body of the message field is an array of 10 transactions from the sensor. The e-miner encrypts each transaction of the message separately for each e-miner using the public key (RSA [48] public-key cryptosystem with a key size of 2048 bits) of each e-miner that participates in the consensus. Finally, the e-miner signs the block, using its RSA public key with a key size of 2048 bits as well. For this experiment, the size of the block is 11644 bytes.

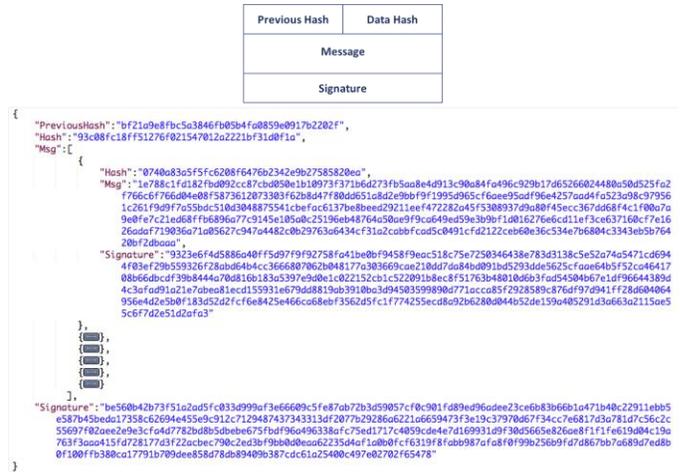

**Figure 7. Structure of a block built by an e-miner.**

Figures 8, 9 and 10 present the results of this experiment. The x-axis represents the number of blocks that the e-miner fabrics and propagates to request consensus approval, which is 100. The y-axis represents the time in seconds that the three e-miners take to achieve consensus.

The graphs show that the time to achieve consensus follows the same pattern in the three scenarios. After the e-miner builds the block, it takes around 2.1 seconds to achieve consensus between the three e-miners.

The interval between transactions does not affect the performance of the three e-miners when collaborating to achieve consensus.

The response times of the consensus component are higher than the response times of the smart-contract component because the consensus component evaluates the attributes of the block and the attributes of each transaction individually.

The e-miner validates the signature of the e-miner that propagated the block. Also, the e-miner validates the hashes of the block. After that, the e-miner decrypts every transaction to verify the signature of the device that initially submitted it and compares the hashes of each transaction. The e-miner also executes the smart-contract component for each decrypted transaction. This approach ensures that before executing the rules of the consensus mechanism, the verification processes of the block and every single transaction are executed by all e-miners that previously accepted to validate that specific group of things.

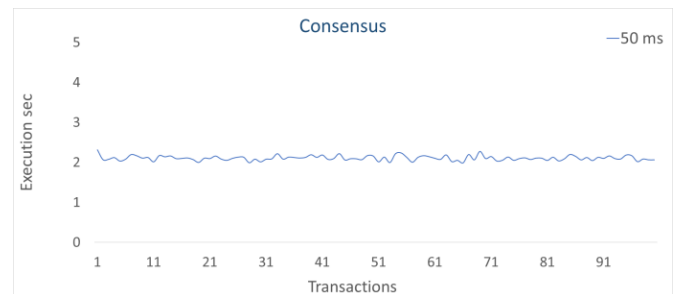

**Figure 8. Evaluation of consensus. Interval of transactions, 50ms.**

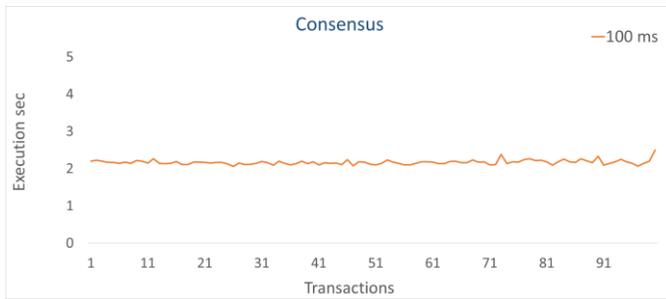

**Figure 9. Evaluation of consensus. Interval of transactions, 100ms.**

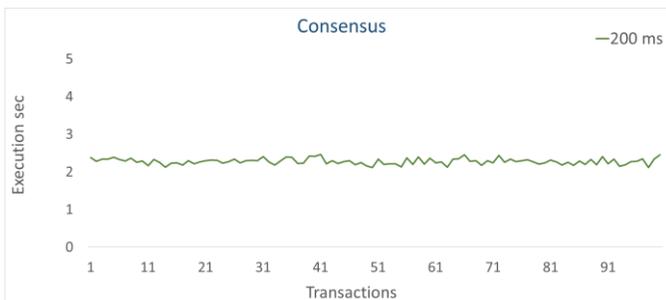

**Figure 10. Evaluation of consensus. Interval of transactions, 200ms.**

### 4.3. Evaluation of in-chain data software-defined component

E-miners keep in-chain metadata of the previously mined blocks in memory to optimize the performance when querying data that does not require the specification of transactions.

When the consensus is achieved, the metadata of the mined block is kept in memory. An in-chain block has three fields, previous hash, data hash, and the signature of the e-miner that mined it.

Additionally, when the consensus is achieved, a file with all the data of the validated block is stored locally. E-miners store a file for each mined block locally in JSON format. The e-miner moves these files to a persistent repository in a Fog node every time 10 blocks achieve consensus. For this experiment in which the block fabric is activated every 10 transactions, the size of each metadata block is 636 bytes.

## 5. Conclusions

This research introduces the novel idea of encapsulating the features of blockchain into software-defined components to distribute them towards edge devices.

This work encapsulates three features of blockchain into software-defined components, smart contract, consensus, and in-chain data. These software-defined blockchain components are customized to fulfill the requirements of IoT networks.

We call each device hosting a software-defined blockchain component an e-miner (edge miner). An e-miner can host any of the blockchain components separately. This approach breaks the barrier of limited computing capabilities at the edge level when integrating blockchain solutions in IoT.

This research uses Edison SoC as e-miners. The results of the experiments show that the characteristics of these e-miners can successfully host the three blockchain components. Additionally, the results of evaluations show that the blockchain components perform well working separated but collaborating and supporting each other to get a blockchain implementation at the edge of the IoT network.

Results of tests may vary due to differences in physical resources, programming languages, or communication protocols.

Overall, this research makes the following contributions to IoT systems. First, this research introduces the idea of building software-defined blockchain components for IoT. Second, enables customization of software-defined blockchain components to deal with the characteristics and requirements of IoT networks. Finally, introduces the definition of e-miners, which are edge devices that host specific software-defined blockchain components.

This research shifts the focus towards edge-centric IoT implementations of cooperative blockchain components. This focus allows us to integrate private blockchain solutions in IoT close to things network.

## 6. References

[1] L. Tan, "Future internet: The Internet of Things," *2010 3rd Int. Conf. Adv. Comput. Theory Eng.*, pp. V5-376-V5-380, 2010.

[2] K. Ashton, "That 'Internet of Things' Thing - RFID Journal," *RFiD J.*, vol. 22, no. 7, pp. 97–114, 2009.

[3] P. N. Howard, "Sketching out the Internet of Things trendline," *Brookings*, 2015. [Online]. Available: https://www.brookings.edu/blog/techtank/2015/06/09/sketching-out-the-internet-of-things-trendline/. [Accessed: 05-Mar-2018].

[4] E. Kaku, R. Orji, J. Pry, K. Sofranko, R. K. Lomotey, and R. Deters, "Privacy Improvement Architecture for IoT," 2018.

[5] J. Gubbi, R. Buyya, S. Marusic, and M. Palaniswami, "Internet of Things (IoT): A vision, architectural elements, and future directions," *Futur. Gener. Comput. Syst.*, vol. 29, no. 7, pp. 1645–1660, Sep. 2013.

[6] F. Bonomi, R. Milito, J. Zhu, and S. Addepalli, "Fog Computing and Its Role in the Internet of Things," *Proc. first Ed. MCC Work. Mob. cloud Comput.*, pp. 13–16, 2012.

[7] R. Cortés, X. Bonnaire, O. Marin, and P. Sens, "Stream Processing of Healthcare Sensor Data: Studying User Traces to Identify Challenges from a Big Data Perspective," *Procedia Comput. Sci.*, vol. 52, pp. 1004–1009, 2015.

[8] Cisco Systems, "Fog Computing and the Internet of Things: Extend the Cloud to Where the Things Are," 2015.

[9] L. M. Vaquero and L. Rodero-Merino, "Finding your Way in the Fog: Towards a Comprehensive Definition of Fog Computing," *ACM SIGCOMM Comput. Commun. Rev.*, vol. 44, no. 5, pp. 27–32, 2014.

[10] F. Bonomi, R. Milito, P. Natarajan, and J. Zhu, "Fog Computing: A Platform for Internet of Things and Analytics," *Big Data Internet Things A Roadmap Smart Environ.*, pp. 169–186, 2014.

[11] M. Samaniego, C. Science, R. Deters, and C. Science, "Zero-

Trust Hierarchical Management in IoT," *2018 IEEE Int. Congr. Internet Things*, pp. 88–95, 2015.

[12] Satyanarayanan Mahadev, "The Emergence of Edge Computing," p. 10, 2017.

[13] B. A. A. Nunes, M. Mendonca, X. N. Nguyen, K. Obraczka, and T. Turletti, "A survey of software-defined networking: Past, present, and future of programmable networks," *IEEE Commun. Surv. Tutorials*, vol. 16, no. 3, pp. 1617–1634, 2014.

[14] Morreale and Anderson, *Software Defined Networking Design and Deployment*. 2012.

[15] S. Nakamoto, "Bitcoin: A Peer-to-Peer Electronic Cash System," p. 9, 2008.

[16] M. Castro and B. Liskov, "Practical Byzantine Fault Tolerance," *Proc. Symp. Oper. Syst. Des. Implement.*, no. February, pp. 1–14, 1999.

[17] A. Singh, P. Goyal, and S. Batra, "An optimized round robin scheduling algorithm for CPU scheduling," *Int. J. Comput. Sci. Eng.*, vol. 02, no. 07, pp. 2383–2385, 2010.

[18] Z. Zheng, S. Xie, H. Dai, X. Chen, and H. Wang, "An Overview of Blockchain Technology: Architecture, Consensus, and Future Trends," *Proc. - 2017 IEEE 6th Int. Congr. Big Data, BigData Congr. 2017*, pp. 557–564, 2017.

[19] V. Buterin, "On Public and Private Blockchains," *Ethereum Blog Crypto Renaiss. salon*, pp. 1–24, 2015.

[20] P. Jayachandran, "The difference between public and private blockchains," *Blockchain Unleashed: IBM Blockchain Blog*, 2017. [Online]. Available: https://www.ibm.com/blogs/blockchain/2017/05/the-difference-between-public-and-private-blockchain/ [Accessed: 11-May-2018].

[21] D. Schwartz, N. Youngs, and A. Britto, "The Ripple protocol consensus algorithm.," in *Ripple Labs Inc White Paper*, 2014, p. 5.

[22] "Hyperledger Fabric – Hyperledger." [Online]. Available: https://www.hyperledger.org/projects/fabric. [Accessed: 17-Mar-2018].

[23] H. Kakavand, N. Kost De Sevres, and B. Chilton, "The Blockchain Revolution: An Analysis of Regulation and Technology Related to Distributed Ledger Technologies," *SSRN Electron. J.*, 2017.

[24] K. Christidis and M. Devetsikiotis, "Blockchains and Smart Contracts for the Internet of Things," *IEEE Access*, vol. 4, pp. 2292–2303, 2016.

[25] "Ethereum Project," 2015. [Online]. Available: https://www.ethereum.org/. [Accessed: 18-Mar-2018].

[26] X. Xingmei, Z. Jing, and W. He, "Research on the basic characteristics, the key technologies, the network architecture and security problems of the Internet of things," in *Proceedings of 2013 3rd International Conference on Computer Science and Network Technology*, 2013, pp. 825–828.

[27] A. Al-Fuqaha, M. Guizani, M. Mohammadi, M. Aledhari, and M. Ayyash, "Internet of Things: A Survey on Enabling Technologies, Protocols and Applications," *IEEE Commun. Surv. Tutorials*, vol. PP, no. 99, pp. 1–1, 2015.

[28] P. K. Sharma, M. Y. Chen, and J. H. Park, "A Software Defined Fog Node Based Distributed Blockchain Cloud Architecture for IoT," *IEEE Access*, vol. 6, pp. 115–124, 2018.

[29] A. Dorri, S. S. Kanhere, R. Jurdak, and P. Gauravaram, "Blockchain for IoT security and privacy: The case study of a smart home," in *2017 IEEE International Conference on Pervasive Computing and Communications Workshops (PerCom Workshops)*, 2017, pp. 618–623.

[30] A. Botta, W. De Donato, V. Persico, and A. Pescape, "On the integration of cloud computing and internet of things," *Proc. - 2014 Int. Conf. Futur. Internet Things Cloud, FiCloud 2014*, pp. 23–30, 2014.

[31] M. Samaniego and R. Deters, "Virtual Resources & Blockchain for Configuration Management in IoT," *J. Ubiquitous Syst. Pervasive Networks*, vol. 9, no. 2, pp. 1–13, 2017.

[32] M. Samaniego and R. Deters, "Internet of Smart Things - IoST: Using Blockchain and CLIPS to Make Things Autonomous," *Proc. - 2017 IEEE 1st Int. Conf. Cogn. Comput. ICCC 2017*, pp. 9–16, 2017.

[33] M. Samaniego and R. Deters, "Supporting IoT Multi-Tenancy on Edge Devices," *Proc. - 2016 IEEE Int. Conf. Internet Things; IEEE Green Comput. Commun. IEEE Cyber, Phys. Soc. Comput. IEEE Smart Data, iThings-GreenCom-CPSCom-Smart Data 2016*, vol. 7, pp. 66–73, 2017.

[34] L. Atzori, A. Iera, and G. Morabito, "The Internet of Things: A survey," *Comput. Networks*, vol. 54, no. 15, pp. 2787–2805, 2010.

[35] N. M. M. K. Chowdhury and R. Boutaba, "A survey of network virtualization," *Comput. Networks*, vol. 54, no. 5, pp. 862–876, 2010.

[36] J. Chen, X. Zheng, and C. Rong, "Survey on software-defined networking," *Lect. Notes Comput. Sci. (including Subser. Lect. Notes Artif. Intell. Lect. Notes Bioinformatics)*, vol. 9106, no. 1, pp. 115–124, 2015.

[37] K. Kirkpatrick, "Software-defined Networking," *Commun. ACM*, vol. 56, no. 9, pp. 58–65, 2013.

[38] S. Nastic, S. Sehic, D. H. Le, H. L. Truong, and S. Dustdar, "Provisioning software-defined IoT cloud systems," *Proc. - 2014 Int. Conf. Futur. Internet Things Cloud, FiCloud 2014*, pp. 288–295, 2014.

[39] A. R. Biswas and R. Giaffreda, "IoT and Cloud Convergence: Opportunities and Challenges," *2014 IEEE World Forum Internet Things*, pp. 375–376, 2014.

[40] "Mapping the Landscape of Decentralized Apps: The Case of Ethereum," no. 2008, p. 60629, 2017.

[41] F. Tschorsch and B. Scheuermann, "Bitcoin and beyond: A technical survey on decentralized digital currencies," *IEEE Commun. Surv. Tutorials*, vol. 18, no. 3, pp. 2084–2123, 2016.

[42] W. Assumptions, "Scheduling: Introduction." [Online]. Available: http://pages.cs.wisc.edu/~remzi/OSTEP/cpu-sched.pdf. [Accessed: 08-Oct-2016].

[43] Z. Shelby, "Constrained RESTful Environments (CoRE) Link Format," 2012.

[44] Golang.org, "The Go Programming Language," 2016. [Online]. Available: https://golang.org/pkg/crypto/rsa/#example_EncryptOAEP. [Accessed: 13-Aug-2017].

[45] INTEL, "The Intel® Edison Module | IoT | Intel® Software," 2017. [Online]. Available: https://software.intel.com/en-us/iot/hardware/edison. [Accessed: 02-Sep-2017].

[46] S. Ingenuity, "SparkFun Electronics," 2009. [Online]. Available: https://www.sparkfun.com/. [Accessed: 02-Sep-2017].

[47] D. Eastlake and P. Jones, "US Secure Hash Algorithm 1 (SHA1)," 2001.

[48] J. Jonsson and B. Kaliski, "Public-Key Cryptography Standards (PKCS) #1: RSA Cryptography Specifications Version 2.1," 2003.